\documentclass[10pt,journal,compsoc]{IEEEtran}
\usepackage{times}
\usepackage{soul}
\usepackage{url}
\usepackage[hidelinks]{hyperref}
\usepackage[utf8]{inputenc}
\usepackage[small]{caption}
\usepackage{graphicx}
\usepackage{amsmath}
\usepackage{amsthm}
\usepackage{booktabs}
\usepackage{algorithm}
\usepackage{algorithmic}
\usepackage{amssymb}
\usepackage{stfloats}
\usepackage{multirow}
\usepackage{color}
\usepackage{subfigure}

\newtheorem{definition}{Definition}
\newtheorem{problem}{Problem}

\ifCLASSOPTIONcompsoc
  \usepackage[nocompress]{cite}
\else
  \usepackage{cite}
\fi
\ifCLASSINFOpdf
\else
\fi
\begin{document}
%
\title{PersonalityGate: A General Plug-and-Play GNN Gate to Enhance Cascade Prediction with Personality Recognition Task}
%
%
%
%

\author{Dengcheng~Yan, Jie~Cao, Yiwen~Zhang, and~Hong~Zhong
\IEEEcompsocitemizethanks{
\IEEEcompsocthanksitem Hong~Zhong is the corresponding author.
\IEEEcompsocthanksitem Dengcheng~Yan, Jie~Cao, Yiwen~Zhang, and~Hong~Zhong was with the School of Computer Science and Technology, Anhui University, Hefei 230026, China.\protect\\
E-mail: yanzhou@ahu.edu.cn}
}

%
%

\markboth{Journal of \LaTeX\ Class Files,~Vol.~14, No.~8, August~2015}%
{Shell \MakeLowercase{\textit{et al.}}: Bare Demo of IEEEtran.cls for Computer Society Journals}
%



\IEEEtitleabstractindextext{%
\begin{abstract}
Cascade prediction estimates the size or the state of a cascade from either microscope or macroscope. It is of paramount importance for understanding the information diffusion process such as the spread of rumors and the propagation of new technologies in social networks. Recently, instead of extracting hand-crafted features or embedding cascade sequences into feature vectors for cascade prediction, graph neural networks (GNNs) are introduced to utilize the network structure which governs the cascade effect. However, these models do not take into account social factors such as personality traits which drive human's participation in the information diffusion process. In this work, we propose a novel multitask framework for enhancing cascade prediction with a personality recognition task. Specially, we design a general plug-and-play GNN gate, named PersonalityGate, to couple into existing GNN-based cascade prediction models to enhance their effectiveness and extract individuals' personality traits jointly. Experimental results on two real-world datasets demonstrate the effectiveness of our proposed framework in enhancing GNN-based cascade prediction models and in predicting individuals' personality traits as well.
\end{abstract}

}

\maketitle

\IEEEdisplaynontitleabstractindextext

%
\IEEEpeerreviewmaketitle

\IEEEraisesectionheading{\section{Introduction}\label{sec:introduction}}

%
%
%
%
\IEEEPARstart{T}{he} prevalence of online social networks such as Twitter and Sina Weibo facilitates the fast diffusion of new technologies, ideas, opinions as well as rumors and viruses, and the diffusion process of which is defined as a cascade \cite{yang2019neural}. Super cascades usually have a huge influence on our society and daily life. For example, huge rumor cascades about epidemics such as COVID-19 would significantly influence public health \cite{pennycook2020fighting} while the fast propagation of a product would bring huge profits to the marketers \cite{chen2010scalable}. Thus, predicting the size, growth, or state of a cascade is of paramount importance for applications in various domains.

\begin{figure}[htbp]
	\centering 
	\includegraphics[scale=0.5]{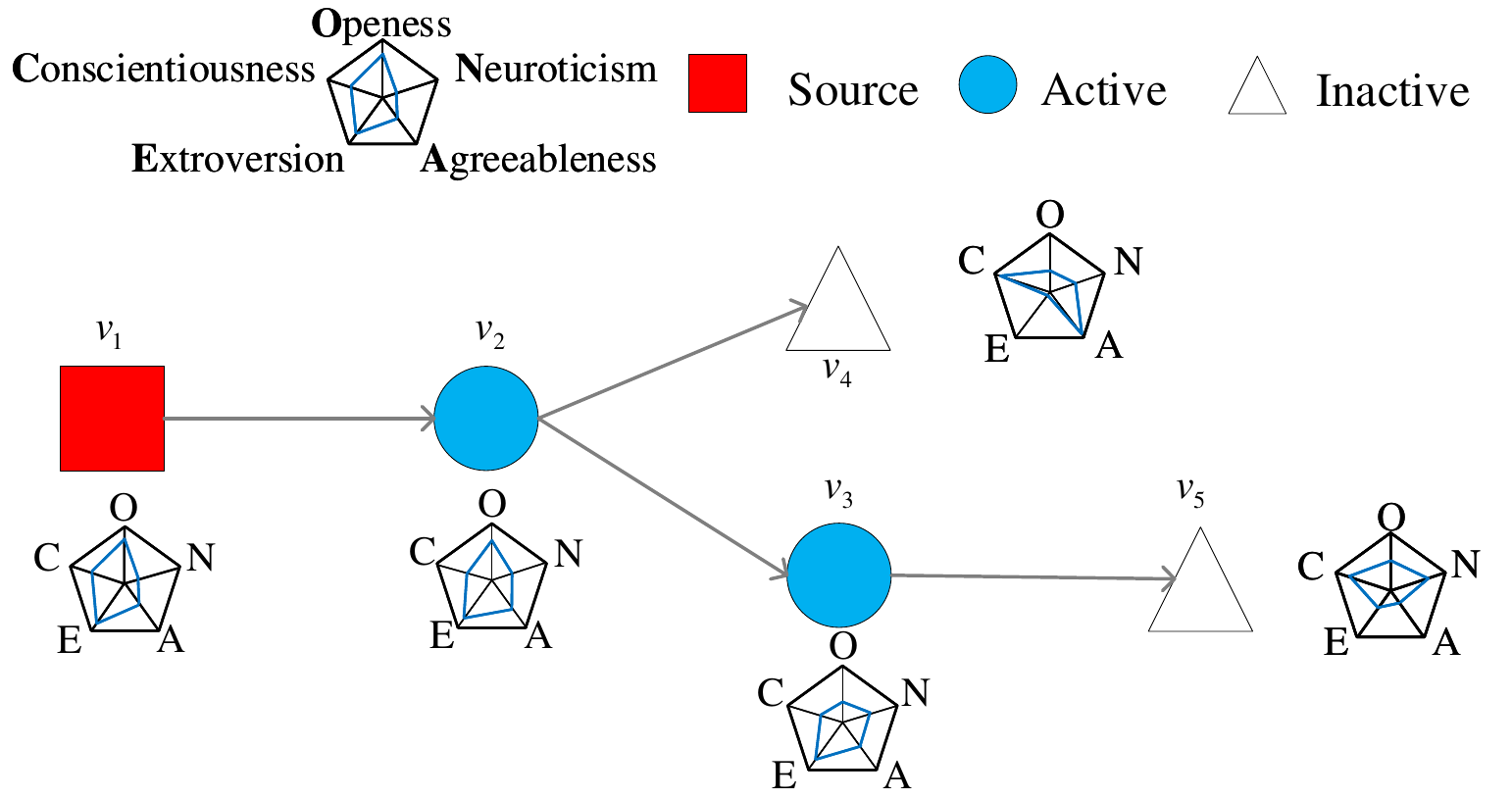}	
	\caption{Motivation example. The effect of individuals' personality traits on information cascade.}
	\label{fig:motivation-example} 
\end{figure}

Recently, cascade prediction is attracting increasing research attentions. Machine learning and deep learning techniques are employed to extract temporal information \cite{li2017deepcas}, content information \cite{liao2019popularity} and network structure information \cite{cao2020coupledgnn} for cascade prediction. However, these statistical learning methods do not take into account the social factors such as personality traits of individuals that drive the formation of cascades, while the mechanism of how a user participates in an information cascade is complicated. Existing psychological and behavioral research has revealed the correlations of individuals' personality traits with information cascade. Users with higher extroversion trait may be more likely to actively share information on a social network \cite{hudson2020bid} while users with high neuroticism trait are more likely to be susceptible to social influence and are less likely to share information on a social network \cite{oyibo2019relationship}. Taking Figure \ref{fig:motivation-example} as an example, in a social network, information from node $v_1$ will be propagated and re-shared by its neighbors $v_2$ and $v_3$ with high extroversion and terminated by its neighbors $v_4$ and $v_5$ with high neuroticism, and the total cascade size is 3. Thus, taking individuals' personality traits into cascade prediction is essential to enhance the performance.

However, existing methods to obtain precise personality traits often face the challenges of both efficiency and effectiveness. On the one hand, traditional questionnaire-based psychological measurement \cite{john1991big,goldberg1993structure} is hardly to be adopted in large-scale online social networks for the reasons of privacy and time cost. While feature-based \cite{tausczik2010liwc} or deep learning-based methods \cite{wei2017beyond} usually need large amounts of user-generated contents such as texts and images. On the other hand, the effectiveness of social behavior-based personality recognition methods \cite{yin2020ffmdetail} relies heavily on handcrafted features from psychological experts.

Considering the tight relationship between individuals' personality traits and their participation in the information cascade process, we utilize the \textbf{M}ultitask approach to \textbf{e}nhance \textbf{CA}scade prediction with \textbf{PE}rsonality recognition task (denoted as \textbf{MeCAPE}). Our main contributions are summarized as follows:

\begin{itemize}
    \item We propose MeCAPE, a novel multitask framework for joint cascade prediction and personality recognition. Individuals' personality traits are revealed from their participation behaviors in the information cascade and the personality-driven mechanism for cascade process is uncovered.
    \item We design a general plug-and-play Graph Neural Network (GNN) gate, PersonalityGate, which simulates the effect of personality traits on the information cascade. It can be easily coupled into various GNN-based cascade prediction models to enhance their effectiveness.
    \item We conduct extensive experiments on two real-world datasets. The results demonstrate the effectiveness of our multitask framework and personality driven GNN gate for both cascade prediction and personality recognition.
\end{itemize}

\begin{figure*}[htp]
	\centering
	\includegraphics[scale=0.205]{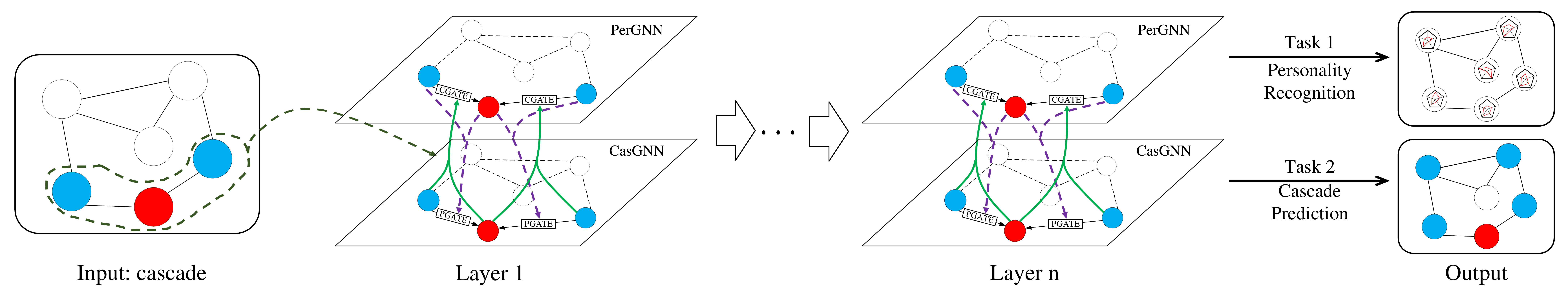}	
	\caption{The framework of MeCAPE.}
	\label{fig:framework} 
\end{figure*}

\section{Preliminaries}

\begin{definition}[Cascade \cite{chen2010scalable}]
A cascade $C_{m}$ is a sequence of activated users $\{u_{1}^{m}, u_{2}^{m}, \cdots, u_{|C_{m}|}^{m}\}$ ($u_{i}^{m} \in U$) participating in the diffusion process of information $m$ ($m \in M$), which is ordered in ascending order according to their activation time. $U$ and $M$ are sets of users and messages, respectively.
\end{definition}

In this work, we only consider the order of activated users in a cascade and ignore the exact timestamps of activation as previous work \cite{yang2019neural} did. 

\begin{problem}[Cascade prediction]
Given a partially observed cascade $C_{m}^{P} = \{u_{1}^{m}, u_{2}^{m}, \cdots, u_{j}^{m}\}$ ($j < |C_{m}|$) of cascade $C_{m}$, cascade size prediction is to predict the total size  $n_{m}$ of cascade $C_{m}$ from the macroscope, and cascade state prediction is to predict the unobserved sequence $C_{m}^{\Delta}$ of cascade $C_{m}$ from microscope.
\end{problem}

In the following sections, we focus on cascade size prediction although cascade state prediction and cascade growth prediction which relate to the activation timestamps are also core directions of cascade prediction. In fact, our proposed MeCAPE framework is also capable of cascade state prediction because it first learns the representation of each node and applies a weighted sum pooling over the representations of all nodes for cascade size prediction. Thus, we can predict cascade state by applying a classifier to the representation of each node to classify it into two classes: in a cascade or not in a cascade.

\begin{definition}[Personality traits \cite{barrick1991bigfive}]
A five dimensional representation of stable and measurable characteristics related to  individual personality, which is denoted as $\boldsymbol{p}_{u} = [p_{O}, p_{C}, p_{E}, p_{A}, p_{N}]$ $(p_{O}, p_{C}, p_{E}, p_{A}, p_{N} \in \mathbb{R})$ for user $u \in U$. 
\end{definition}

Personality recognition is to predict the five dimensions of personality traits, i.e., \textbf{O}penness, \textbf{C}onscientiousness, \textbf{E}xtroversion, \textbf{A}greeableness and \textbf{N}euroticism, which are related to different individual characteristics \cite{yin2020ffmdetail}.

\begin{itemize}
    \item \textbf{Openness}. The traits of being imaginative, curious, broad-minded and tolerant of new ideas.
    \item \textbf{Conscientiousness}. The traits of being careful, responsible, organized and goal-directed.
    \item \textbf{Extroversion}. The traits of being social, active, outgoing and ambitious.
    \item \textbf{Agreeableness}. The traits of being kind, considerate, trusting and cooperative.
    \item \textbf{Neuroticism}. The traits of being anxious, depressed, angry and insecure.
\end{itemize}

\section{Proposed Framework}

To simultaneously reveal individuals' personality traits from their behaviors in information cascade and in turn enhance cascade prediction with personality traits, we propose \textbf{MeCAPE}, a \textbf{M}ultitask framework for \textbf{e}nhancing \textbf{CA}scade prediction with \textbf{PE}rsonality recognition task with a general plug-and-play GNN gate, \textbf{PersonalityGate}.

The overview of the MeCAPE framework is illustrated in Figure \ref{fig:framework}. It models the interplay between individuals' personality traits and the information cascade process and enhances both personality recognition and cascade prediction by coupling PersonalityGate to any existing GNN-based cascade prediction model (denoted as CasGNN hereafter). The CasGNN simulates the formation of an information cascade across the social network that the activation of a user will trigger its neighbors in a continuous manner. The PersonalityGate captures individuals' personality traits from their participation behaviors in a cascade modeled by the CasGNN and in turn provides personality traits driven attention for neighbor influence in the CasGNN.

\subsection{GNN-based Cascade Prediction}
\label{sec:gnn-cas-pred}

GNN-based cascade prediction models utilize a GNN to first obtain a low-dimensional vector representation for each node in a social network by neighborhood information aggregation which captures the influence between node pairs during information cascade. Then a regression model is employed with the representations of the early adopters of a cascade for cascade size prediction.

Generally, given a network $G=(V, E)$ ($V$ and $E$ are set of nodes and edges, respectively) with nodes' features $\boldsymbol{X} \in \mathbb{R}^{|V| \times F}$ ($F$ is the dimension of nodes' feature), GNNs utilize the network structure and node features to learn a representation $\boldsymbol{h}_{v}$ of each node $v$ with a neighborhood aggregation strategy \cite{xu2018gnngeneral}. The neighborhood aggregation runs iteratively and captures the structural information within multi-hop neighborhood. The general form of GNN can be formulated as Equation (\ref{eq:gnn-general-agg}) and (\ref{eq:gnn-general-comb}),

\begin{equation}
    \label{eq:gnn-general-agg}
    \boldsymbol{a}_{v}^{k}=\operatorname{AGG}^{k} \left(\left\{\boldsymbol{h}_{u}^{k}: u \in \mathcal{N}(v)\right\}\right)
\end{equation}

\begin{equation}
    \label{eq:gnn-general-comb}
    \boldsymbol{h}_{v}^{k+1}=\operatorname{COMBINE}^{k} \left(\boldsymbol{h}_{v}^{k}, \boldsymbol{a}_{v}^{k}\right)
\end{equation}
where $\boldsymbol{h}_{u}^{k}$ and $\mathcal{N}(v)$ are the representation vector of node $u$ at the $k$-th layer and the neighbor nodes set of node $v$. $\operatorname{AGG}^{k}\left(\cdot\right)$ and $\operatorname{COMBINE}^{k}\left(\cdot\right)$ are the functions for aggregating and combing neighborhood information at the $k$-th layer.

Then the cascade prediction model $\operatorname{{CASCADE}\left(\cdot\right)}$ uses the representations of all early adopters in the partially observed cascade $C_{m}^{P}$ output by the last layer of $\operatorname{{GNN}\left(\cdot\right)}$ to predict cascade size $n_{m}$ as Equation (\ref{eq:cas-size-pred-general}).

\begin{equation}
    \label{eq:cas-size-pred-general}
    n_{m}=\operatorname{CASCADE} \left( \operatorname{GNN} \left( C_{m}^{P} \right) \right)
\end{equation}

\subsection{PersonalityGate}

Considering the influence of individual personality traits on the cascade process, a personality graph neural network (PerGNN) is introduced and coupled into the cascade graph neural network (CasGNN) described in Section \ref{sec:gnn-cas-pred} with PersonalityGate.

The PerGNN captures individuals' participation behaviors of information cascade from CasGNN to reveal their personality traits, which is illustrated in Equation (\ref{eq:personalitygate-per-agg}) and (\ref{eq:personalitygate-per-comb}).

\begin{equation}
    \label{eq:personalitygate-per-agg}
    \boldsymbol{a}_{v}^{k}=\operatorname{AGG}_{P}^{k} \left(\left\{ \operatorname{CGATE}^{k} \left( \boldsymbol{c}_{u}^{k},\boldsymbol{c}_{v}^{k} \right) \boldsymbol{p}_{u}^{k}: u \in \mathcal{N}(v)\right\}\right)
\end{equation}

\begin{equation}
    \label{eq:personalitygate-per-comb}
    \boldsymbol{p}_{v}^{k+1}=\operatorname{COMBINE}_{P}^{k} \left(\boldsymbol{p}_{v}^{k}, \boldsymbol{a}_{v}^{k}\right)
\end{equation}
where $\mathcal{N}(v)$ is the neighborhood of user $v$, $\boldsymbol{c}_{u}^{k} \in \mathbb{R}^{d_c}$ and $\boldsymbol{p}_{u}^{k} \in \mathbb{R}^{d_p}$ are the representations from the $k$-th layer of CasGNN and PerGNN, respectively. ${d_c}$ and ${d_p}$ are their representation dimensions. $\operatorname{AGG}_{P}^{k}\left(\cdot\right)$ and $\operatorname{COMBINE}_{P}^{k}\left(\cdot\right)$ are functions for aggregating and combining neighbors' information in the $k$-th layer of PerGNN, respectively. $\operatorname{CGATE^{k} \left(\cdot\right)}$ is a gating mechanism at $k$ layer for capturing an individual's participation behaviors of the information cascade.

The CasGNN utilizes the information of individuals' personality traits from PerGNN to guide the cascade process, which is illustrated in Equation (\ref{eq:personalitygate-cas-agg}) and (\ref{eq:personalitygate-cas-comb}). 

\begin{equation}
    \label{eq:personalitygate-cas-agg}
    \boldsymbol{b}_{v}^{k}= \operatorname{AGG}_{C}^{k} \left(\left\{ \operatorname{PGATE}^{k} \left( \boldsymbol{p}_{u}^{k}, \boldsymbol{p}_{v}^{k} \right) \boldsymbol{c}_{u}^{k}: u \in \mathcal{N}(v)\right\}\right)
\end{equation}

\begin{equation}
    \label{eq:personalitygate-cas-comb}
    \boldsymbol{c}_{v}^{k+1}=\operatorname{COMBINE}_{C}^{k} \left(\boldsymbol{c}_{v}^{k}, \boldsymbol{b}_{v}^{k}\right)
\end{equation}
where $\operatorname{AGG}_{C}^{k}\left(\cdot\right)$ and $\operatorname{COMBINE}_{C}^{k}\left(\cdot\right)$ are functions for aggregating and combining neighbors' information in the $k$-th layer of CasGNN, respectively. $\operatorname{PGATE}^{k} \left(\cdot\right)$ indicates the PersonalityGate, which is a gating mechanism  at $k$-th layer for capturing the effect of personality traits on cascading effect.

In this work, for simplicity, we implement $\operatorname{PGATE}^{k}$ and $\operatorname{CGATE}^{k}$ as a weighted sum of the weighted concatenation of node pairs, as shown in Equation (\ref{eq:casgate-impl}) and (\ref{eq:personalitygate-impl}),

\begin{equation}
    \label{eq:casgate-impl}
    \operatorname{CGATE}^{k} \left( \boldsymbol{c}_{u}^{k},\boldsymbol{c}_{v}^{k} \right)= \boldsymbol{\beta}_{CG}^{k} \left(\boldsymbol{W}_{CG}^{k} \boldsymbol{c}_{u}^{k}|| \boldsymbol{W}_{CG}^{k} \boldsymbol{c}_{v}^{k}\right)
\end{equation}

\begin{equation}
    \label{eq:personalitygate-impl}
    \operatorname{PGATE}^{k} \left( \boldsymbol{p}_{u}^{k},\boldsymbol{p}_{v}^{k} \right)= \boldsymbol{\beta}_{PG}^{k} \left(\boldsymbol{W}_{PG}^{k} \boldsymbol{p}_{u}^{k}||\boldsymbol{W}_{PG}^{k} \boldsymbol{p}_{v}^{k}\right)
\end{equation}
where $\boldsymbol{W}_{CG}^{k} \in \mathbb{R}^{d_{c}^{k+1} \times d_{c}^{k}}$, $\boldsymbol{W}_{PG}^{k} \in \mathbb{R}^{d_{p}^{k+1} \times d_{p}^{k}}$ are weight matrices to transform the representation from dimension $d_{c}^{k}$ to $d_{c}^{k+1}$, and from dimension $d_{p}^{k}$ to $d_{p}^{k+1}$, respectively. The $\boldsymbol{\beta}_{CG}^{k} \in \mathbb{R}^{2d_{c}^{k+1}}$, $\boldsymbol{\beta}_{PG}^{k} \in \mathbb{R}^{2d_{p}^{k+1}}$ are weight vectors. However, it is worth noting that any form that combines the representations of node pairs is applicable.

In this work, we will plug PersonalityGate into GCN, GAT and CoupledGNN (denoted as GCN-p, GAT-p and CoupledGNN-p, respectively). The $\operatorname{AGG}\left(\cdot\right)$ and $\operatorname{COMBINE}\left(\cdot\right)$ of both CasGNN and PerGNN for GCN, GAT and CoupledGNN employ weighted sum pooling strategy  and element-wise mean pooling strategy, respectively. An example of GCN-p implementation is shown in Equation (\ref{eq:detail-struct-coupled-gcn-per}) and (\ref{eq:detail-struct-coupled-gcn-cas}), and the implementations of GAT-p and CoupledGNN-p are similar to GCN-p.

\begin{equation}
    \label{eq:detail-struct-coupled-gcn-per}
    \boldsymbol{p}_{v}^{k+1}=\sigma \left( \boldsymbol{W}_{P}^{k} \boldsymbol{p}_{v}^{k} + \boldsymbol{W}_{P}^{k} \sum_{u \in \mathcal{N}(v)}  \operatorname{CGATE}\left( \boldsymbol{c}_{u}^{k},\boldsymbol{c}_{v}^{k} \right)\boldsymbol{p}_{u}^{k}\right)
\end{equation}

\begin{equation}
    \label{eq:detail-struct-coupled-gcn-cas}
    \boldsymbol{c}_{v}^{k+1}=\sigma \left( \boldsymbol{W}_{C}^{k} \boldsymbol{c}_{v}^{k} + \boldsymbol{W}_{C}^{k} \sum_{u \in \mathcal{N}(v)}  \operatorname{PGATE}\left( \boldsymbol{p}_{u}^{k},\boldsymbol{p}_{v}^{k} \right)\boldsymbol{c}_{u}^{k}\right)
\end{equation}
where $\boldsymbol{W}_{C}^{k} \in \mathbb{R}^{d_{c}^{k+1} \times d_{c}^{k}}$, $\boldsymbol{W}_{P}^{k} \in \mathbb{R}^{d_{p}^{k+1} \times d_{p}^{k}}$ are weight matrices to transform the representation from dimension $d_{c}^{k}$ to $d_{c}^{k+1}$, and from dimension $d_{p}^{k}$ to $d_{p}^{k+1}$. $\sigma$ is a nonlinear activation function.

\subsection{Multitask Prediction}

After $K$ rounds of interplays between PerGNN and CasGNN, the representations of personality traits and cascade status are obtained from PerGNN and CasGNN, respectively, i.e., $\boldsymbol{p}_{u}^{K}$ and $\boldsymbol{c}_{u}^{K}$. Then specific regression models can be employed and chained after these representations for corresponding tasks, i.e., personality recognition and cascade prediction. In our proposed MeCAPE framework, MLP is chosen as the regression model for simplicity, but in fact any kind of regression models is applicable.

\textbf{Cascade prediction task} predicts the size of affected users for each cascade information. The predicted size $\hat{n}_{m}$ of a partially observed cascade $C_{m}^P$ is defined as Equation (\ref{eq:cas-pred}),

\begin{equation}
    \label{eq:cas-pred}
    \hat{n}_{m} = \sum_{u \in V} \sigma \left(\boldsymbol{W}_{CP}\boldsymbol{c}_{u}^{K}\right)
\end{equation}
where $\boldsymbol{W}_{CP} \in \mathbb{R}^{d_{c}^{K}}$ is the weight vector, $\sigma$ is the sigmoid function.

Specially, CoupledGNN introduces a StateGate to capture individuals' activation state $s_{u}^{K} \in \mathbb{R}$, thus we simply sum up the status of all users for each cascade $C_{m}$, as illustrated in Equation (\ref{eq:cas-pred-coupledgnn}).

\begin{equation}
    \label{eq:cas-pred-coupledgnn}
    \hat{n}_{m} = \sum_{u \in V} s_{u}^{K}
\end{equation}

The optimization objective for cascade prediction task is to minimize the mean relative square error (MRSE) between the predicted cascade size $\hat{n}_{m}$ and ground truth cascade size $n_{m}$ ,as shown in Equation (\ref{eq:loss-caspred}).

\begin{equation}
    \label{eq:loss-caspred}
    L_{CAS} = \frac{1}{M} \sum_{m=1}^{M} \left( \frac{\hat{n}_{m}-n_{m}}{n_{m}} \right)^{2}
\end{equation}

\textbf{Personality recognition task} predicts the score of each trait of BigFive model for each individual. The predicted value $\hat{\boldsymbol{q}}_{u} \in \mathbb{R}^5$ is defined in Equation (\ref{eq:per-pred}).

\begin{equation}
    \label{eq:per-pred}
    \hat{\boldsymbol{q}}_{u} = \text{ReLU} \left(\boldsymbol{W}_{PP}\boldsymbol{p}_{u}^{K}\right)
\end{equation}
where $\boldsymbol{W}_{PP} \in \mathbb{R}^{5 \times d_{p}^{K}}$ is weight matrix to transform the representation from dimension $d_{p}^{K}$ to $5$.

Same as cascade prediction task, the optimization objective for the personality recognition task is to minimize the mean relative square error between predicted personality traits $\hat{\boldsymbol{q}}_{u}$ and the ground truth personality traits $\boldsymbol{q}_{u}$, as illustrated in Equation (\ref{eq:loss-perpred}).

\begin{equation}
\label{eq:loss-perpred}
L_{P E R}=\frac{1}{V} \sum_{v \in V}\left\|(\hat{\boldsymbol{q}}_{v}-\boldsymbol{q}_{v})\otimes(\frac{1}{\boldsymbol{q}_{v}})^{T}\right\|_{2}^{2}
\end{equation}
where $\otimes$ is Hadamard product, that is element-wise product.
Finally, the overall optimization objective for the MeCAPE framework is a combination of both personality recognition and cascade prediction which is balanced with a parameter $\lambda$, as shown in Equation (\ref{eq:loss-total}).

\begin{equation}
    \label{eq:loss-total}
    L = L_{CAS} + \lambda L_{PER}
\end{equation}

\section{Experiments}

We evaluate the performance and effectiveness of our proposed framework on two real-world datasets, Sina Weibo and DBLP, against multiple baselines.

\subsection{Datasets}

\paragraph{Sina Weibo.} Sina Weibo is a popular online social network platform in China, where users can post, repost and comment on a microblog, and establish follow relations with others. The Sina Weibo dataset from previous works \cite{zhang2013social} consists of 1.78 million users and 308 million user relationships as well as cascade details of 300000 popular microblogs. We extract a subset of the dataset following the ideas in \cite{cao2020coupledgnn} and finally get 2454 users with 139591 relations as well as 99 pieces of information cascade. The ground truth personality traits of each user are obtained from the API of Personality Insights \footnote{https://personality-insights-demo.ng.bluemix.net/} with all the microblogs of the user.

\paragraph{DBLP.} DBLP is a computer science bibliography platform. The DBLP dataset is a citation network released by AMiner \cite{tang2008arnetminer}. The citation of a paper among authors is treated as a cascade. We randomly sample 743 articles published in 2003 and select the 101 papers with citations over 10. We construct a citation network among the 2015 authors related to these articles by adding 55433 edges from authors to authors being cited. The ground truth personality traits of each author are obtained from Personality Insights with all the abstracts of the articles written by the author. The ground truth personality traits for DBLP are obtained from the webpage of Personality Insights and the personality traits are percentiles.

\subsection{Baselines}

In the experiment, we evaluate the effectiveness of our proposed MeCAPE framework and PersonalityGate. PersonalityGate is plugged into three classic graph neural networks, i.e., graph convolutional network (GCN) \cite{kipf2017gcn}, graph attention network (GAT) \cite{velivckovic2017gat} and CoupledGNN \cite{cao2020coupledgnn} and their corresponding PersonalityGate-plugged GNNs are denoted as GCN-p, GAT-p, and CoupledGNN-p, respectively.

\begin{table}[!htp]
	\centering
	\begin{tabular}{lrrrr}
		\toprule
		& \multicolumn{2}{c}{Sina Weibo} & \multicolumn{2}{c}{DBLP} \\
		& RMRSE          & MAPE          & RMRSE  & MAPE \\
		\midrule
		FBC-r               & 0.4110          & 0.3073          & 0.5540          & 0.4949 \\
		FBC-m               & 0.3618          & 0.2881          & 0.5665          & 0.5008 \\
		DeepCas             & 0.3862          & 0.2876          & 1.4484          & 1.0729 \\
		\midrule
		GCN                 & 0.2799          & 0.2315          & 0.6478          & 0.5868 \\
		GCN-p               & \textbf{0.2780} & \textbf{0.2299} & \textbf{0.6432} & \textbf{0.5784} \\
		GAT                 & 0.2230          & 0.1821          & 0.6477          & 0.5866 \\
		GAT-p               & \textbf{0.2105} & \textbf{0.1728} & \textbf{0.6372} & \textbf{0.5625} \\
		CoupledGNN          & 0.3272          & 0.2971          & 0.4882          & 0.4657 \\
		CoupledGNN-p        & \textbf{0.2361} & \textbf{0.2212} & \textbf{0.3732} & \textbf{0.3579} \\
		\bottomrule           
	\end{tabular}
	\caption{The overall performance of cascade prediction task.}
	\label{tab:performance-cas}
\end{table}

For the cascade prediction task, PersonalityGate-plugged GNNs are first compared with their originals to evaluate the effectiveness of PersonalityGate in enhancing cascade prediction. The performance of PersonalityGate-plugged GNNs in cascade prediction are also compared with classic cascade prediction models, i.e., feature-based regression (FBC-r), feature-based multilayer perceptron (FBC-m) and DeepCas \cite{li2017deepcas}. For the personality recognition task, GCN-p, GAT-p and CoupledGNN-p are compared with classic methods including feature-based regression (FBP-r), feature-based multilayer perceptron (FBP-m), PersonalityRecognizer (PR) \cite{mairesse2007personalityrecognizer} and LIWC \cite{schwartz2013personality}.

\subsection{Evaluation Metrics}

In the experiment, we evaluate the performance of our proposed framework and baselines using two evaluation metrics Root Mean Relative Square Error (RMRSE) and Mean Absolute Percentage Error (MAPE) \cite{cao2020coupledgnn} defined in Equation (\ref{eq:eval-metrics-rmrse}) and (\ref{eq:eval-metrics-mape}), respectively.

\begin{equation}
    \label{eq:eval-metrics-rmrse}
	RMRSE=\sqrt{\frac{1}{N} \sum_{n=1}^{N}\left(\frac{y_{i} - \hat{y}_{i}}{y_{i}}\right)^{2}}
\end{equation}

\begin{equation}
    \label{eq:eval-metrics-mape}
	MAPE=\frac{1}{N} \sum_{i=1}^{N}\left|\frac{\hat{y}_{i}-y_{i}}{y_{i}}\right|
\end{equation}
where $\hat{y}_{i}$ and $y_{i}$ denote the predicted value and ground-truth value for record $i$, respectively, and $N$ is the total number of records. It is worth noting that for both evaluation metrics the smaller value indicates better performance.

\subsection{Experimental Settings}

For feature-based baselines, user-related and cascade-related features are extracted and used for personality recognition and cascade prediction, respectively.

User-related features for the Sina Weibo dataset include the average length of a user's posts, the number of @ appears in a user's posts, whether the majority of a user's posts are published on weekdays, whether the majority of a user's posts are published during the daytime (8:00-18:00), and the social network structure measures (the coreness, pagerank, hub score, authority score, eigenvector centrality, and clustering coefficient). User-related features for the DBLP dataset include the average length of the abstracts of  an author's published papers, the year with most published papers and the same network structure measures as for the Sina Weibo dataset.

\begin{table}[ht]
	\centering
	\begin{tabular}{lrrrr}
		\toprule
		& \multicolumn{2}{c}{Sina Weibo} & \multicolumn{2}{c}{DBLP} \\
		& RMRSE          & MAPE          & RMRSE  & MAPE \\
		\midrule
		LIWC                  & 0.3797           & 0.3171           & -               & - \\
		PR & 0.2241           & 0.1685           & -               & - \\
		FBP-r                 & 0.1722           & 0.1292           & 3.3252          & 1.4660 \\
		FBP-m                 & 0.1215           & 0.0778           & 2.3008          & 1.2255 \\
		\midrule
		GCN-p                 & \textbf{0.1108}           & 0.0709           & 0.6730          & 0.5084 \\
		GAT-p                 & 0.1127           & \textbf{0.0681}  & 0.6808          & 0.5251 \\
		CoupledGNN-p          & 0.1210  & 0.0722           & \textbf{0.6236} & \textbf{0.4836} \\
		\bottomrule           
	\end{tabular}
	\caption{The overall performance of personality recognition task.}
	\label{tab:performance-per}
\end{table}

Cascade-related features for the Sina Weibo dataset include the content length of the message in cascade, the number of @ appears in a message in cascade, whether the message in cascade is posted on weekdays, whether the message in cascade is posted during the daytime, the size of the cascade and the average of each network structure measure of all users in a cascade. Cascade-related features for the DBLP dataset include the abstract length of a paper, the published year of a paper, the number of papers that cite a paper and the average of each network structure measure of the authors that cite a paper.

The number of layers of MLP for FBC-m and FBP-m are 3. The dimension of the embeddings $d$ of CoupledGNN and CoupledGNN-p are 38 (32 random embeddings plus 6 network structure features, i.e., coreness, pagerank, hub score, authority score, eigenvector centrality and clustering coefficient). The dimension of the embeddings $d$ of GCN, GAT, GCN-p and GAT-p are 39 (the previous 38 dimensions are the same as CoupledGNN and the last dimension indicates whether the user is in a cascade).
The learning rate is 0.0005. The coefficient $\lambda$ is 10.0 in Sina Weibo, 0.001 in the DBLP dataset. The number of GNN layers K is 3.

\subsection{Experimental Results}

\paragraph{Cascade Prediction.} As shown in Table \ref{tab:performance-cas}, compared with conventional GNN-based cascade prediction models (GCN, GAT and CoupledGNN), the corresponding PersonalityGate-plugged-in versions (GCN-p, GAT-p and CoupledGNN-p) have obvious improvement, indicating the effectiveness and strong universality of the personality gating mechanism introduced by PersonalityGate in enhancing GNN-based cascade prediction models. Specially, GAT-p and CoupledGNN-p have more significant improvement than GCN-p because the former two models both introduce attention mechanism. Moreover, GNN-based and PersonalityGate-plugged-in GNN-based cascade prediction models all outperform conventional methods, indicating the necessity of introducing network structure.

\paragraph{Personality Recognition.} As shown in Table \ref{tab:performance-per}, the results indicate our method outperforms the conventional personality recognition methods that usually reveal users' personality traits from their generated contents (LIWC, PR) or by handcrafted features of their behaviors (FBP-r, FBP-m). Our proposed PersonalityGate can automatically aggregate information from only users' cascading behaviors for personality recognition, indicating its wide range of applications. On the one hand, because PersonalityGate is plugged into GNN-based cascade prediction models to aggregate information from cascade sequence, its personality recognition does not need additional content information. On the other hand, PersonalityGate employs a GNN structure thus extracts better features than handcrafted features. It is worth noticing that LIWC and PR are not applicable to DBLP, for the ground-truth personality is fetched from the web page of Personality Insights and is in a style of percentile while LIWC and PR predict raw scores.

\begin{figure}
	\centering
	\subfigtopskip=0pt
	\subfigure[Sina Weibo]{
		\includegraphics[scale=0.25]{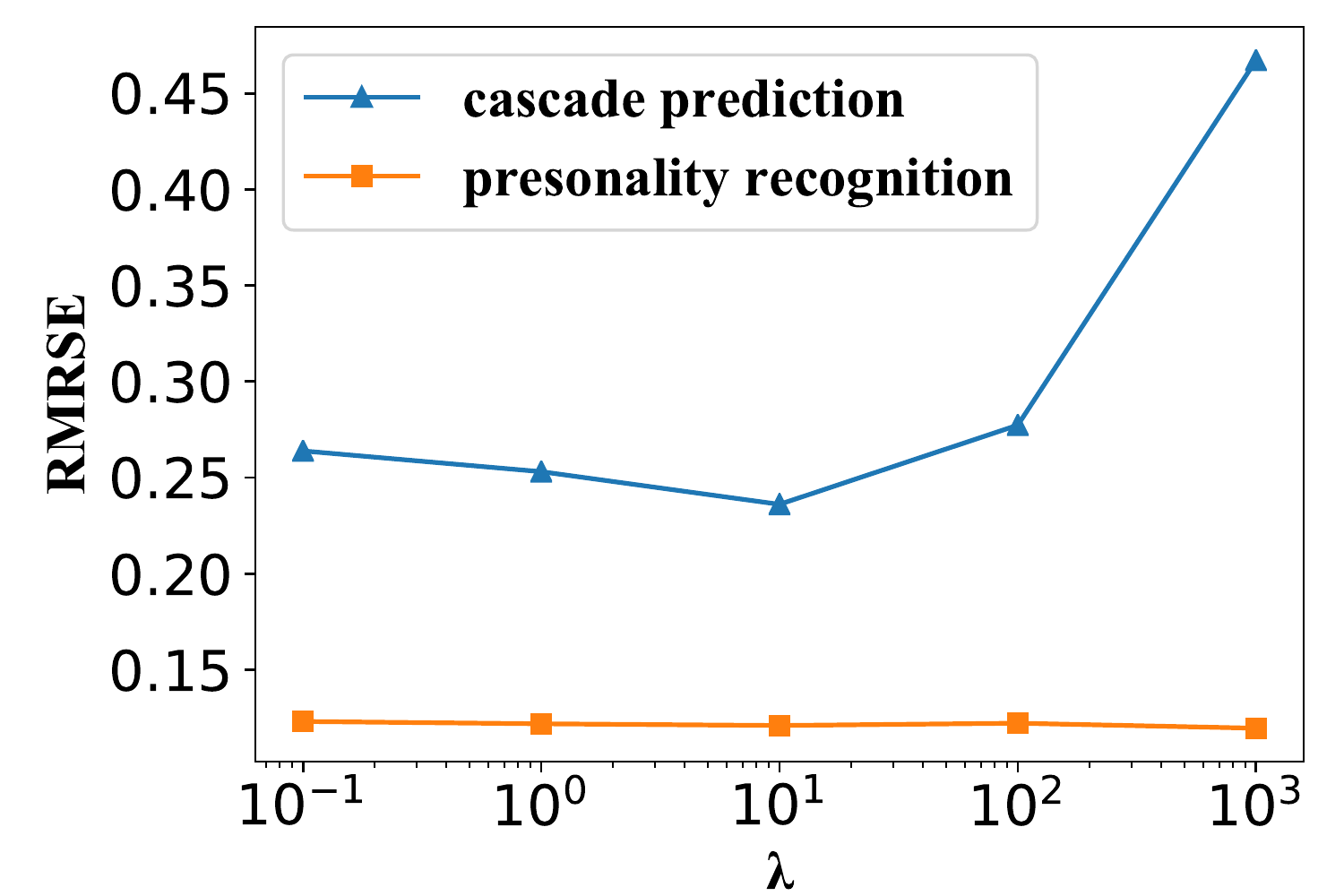}
		\label{fig:effect-lambda-rmrse-weibo}
	}
	\subfigure[Sina Weibo]{
		\includegraphics[scale=0.25]{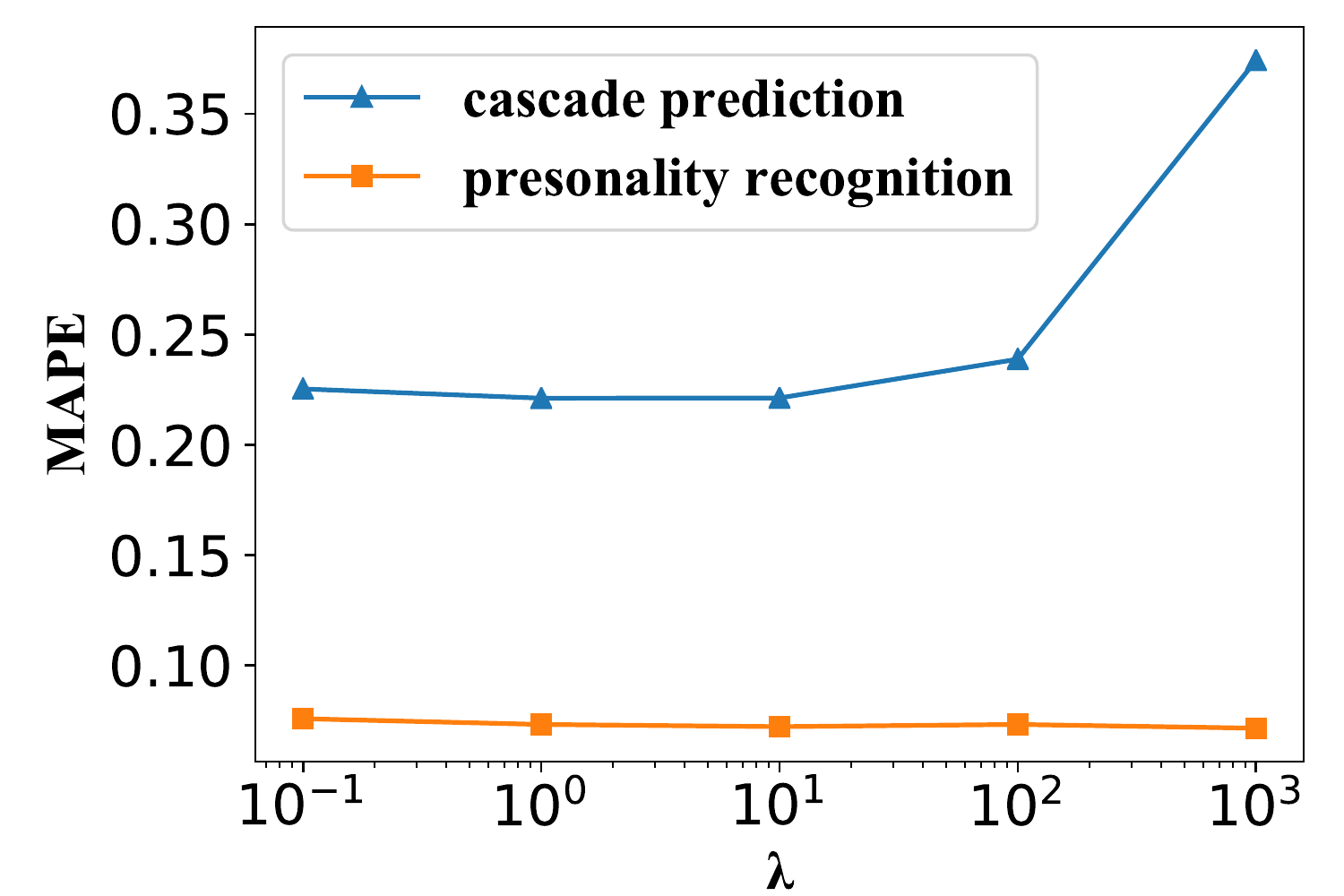}
		\label{fig:effect-lambda-mape-weibo}
	}
	\subfigure[DBLP]{
		\includegraphics[scale=0.25]{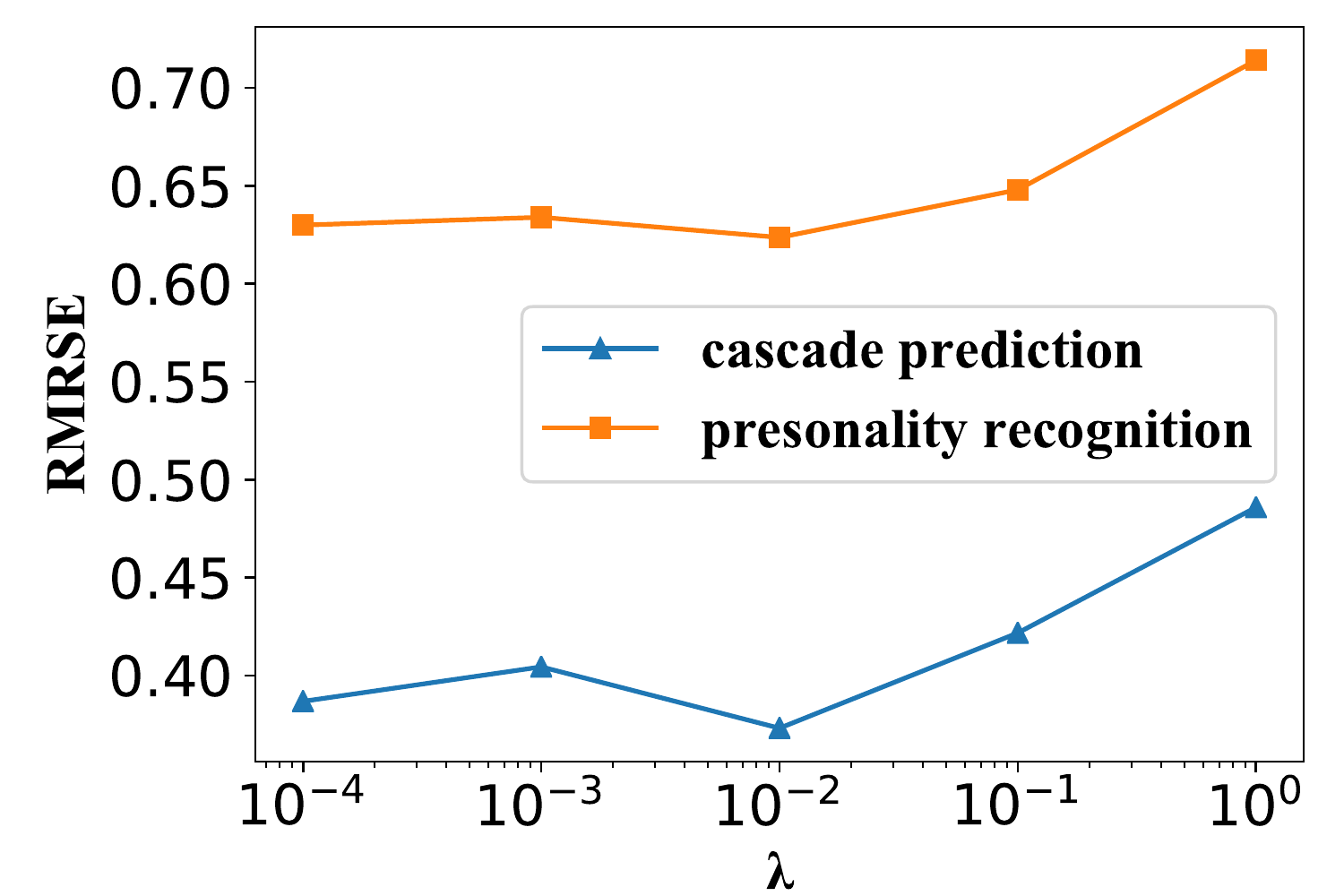}
		\label{fig:effect-lambda-rmrse-dblp}
	}
	\subfigure[DBLP]{
		\includegraphics[scale=0.25]{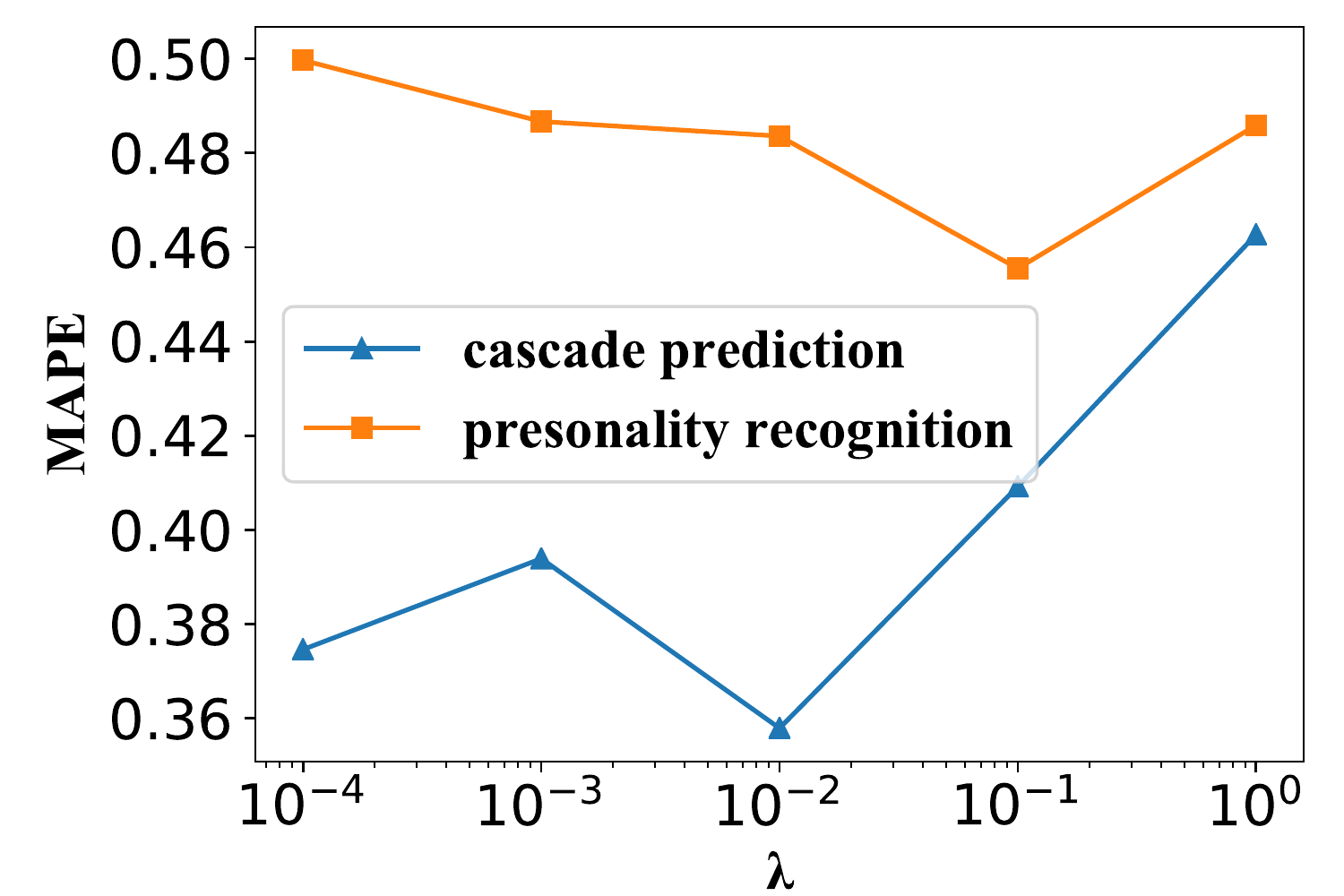}
		\label{fig:figs/effect-lambda-mape-dblp}
	}
	\caption{The effect of $\lambda$.}
	\label{fig:effect-lambda}
\end{figure}

\paragraph{The effect of personality recognition weight $\lambda$.} Our proposed MeCAPE framework is a multitask framework which utilizes personality recognition task to enhance cascade prediction task. As shown in Figure \ref{fig:effect-lambda}, we can find that with the importance increment of the personality recognition task, i.e., with the increment of $\lambda$, the RMRSE and MAPE on both datasets first decrease and then increase, indicating the effectiveness of the personality recognition task in enhancing cascade prediction task. Too small weight for personality recognition task would not integrate enough personality traits information to enhance cascade prediction. Too large weight for personality recognition task would overvalue the driven force of personality traits in the information cascade process and undervalue other factors.

\section{Related Work}

\paragraph{GNN-based cascade prediction.} A cascade is the diffusion process of information and cascade prediction aims to estimate either the size or growth of a cascade from the macroscope \cite{li2017deepcas} or the state of each user in a cascade from the microscope \cite{yang2019neural}. Recently, machine learning and deep learning techniques are employed to extract temporal information \cite{li2017deepcas}, content information \cite{liao2019popularity},  and network structure information \cite{cao2020coupledgnn} for cascade prediction. Specially, CoupledGNN \cite{cao2020coupledgnn} utilizes two layers of graph neural networks (GNNs), a kind of deep learning technique on graphs \cite{xu2018gnngeneral} for cascade size prediction and compares itself with methods using graph convolutional network (GCN) \cite{kipf2017gcn} and graph attention network (GAT) \cite{velivckovic2017gat}.

\paragraph{Personality recognition.} Personality is the stable and unique characteristic of individuals which relates to individuals' behavioral, temperamental, emotional, and mental states \cite{kaushal2018emerging}. Personality recognition aims to measure the multiple characteristics of individuals, i.e., personality traits. Five Factor Model \cite{barrick1991bigfive}, i.e., Big Five, which is the most popular measure for personality traits, classify personality traits into five dimensions: openness, conscientiousness, extroversion, agreeableness, and neuroticism. Traditional psychological researches focus on designing questionnaires to calculate the scores of personality traits \cite{john1991big,goldberg1993structure}. Linguistics-based methods extract expert defined linguistic features from user-generated texts and employ statistical methods such as regression to predict the scores of personality traits \cite{tausczik2010liwc}. Recently, deep learning \cite{wei2017beyond} is utilized to automatically extract features from various kinds of user-generated data, including texts and images.

\paragraph{Multitask learning.} Multitask learning \cite{caruana1997multitask}  simultaneously learns multiple related tasks instead of one task at a time of conventional machine learning. Multitask learning is based on the assumption that the data distributions of related tasks are similar and uses shared representations \cite{evgeniou2004regularized} or parameter constraints \cite{gong2012robust} across tasks to enhance the performance and improve the generalization of each task. Recently, with the prevalence of graph data mining and graph neural networks, the general ideas of multitask learning such as shared representations and parameter constraints are introduced to graphs \cite{nassif2020multitask} with various applications such as graph classification \cite{xie2020multi}. 

\section{Conclusion and Future Work}

In this work, we study the problem of enhancing the GNN-based cascade prediction model with a personality recognition task from a multitask learning approach. The general idea behind this approach is inspired by the views from psychological and behavioral researches that individuals' personality traits and their participation in the information cascade process are tightly coupled. Individuals' personality traits can be revealed from their participation behaviors in the information cascade and individuals' personality traits drive their participation in information cascade process. Specially, we design a general plug-and-play GNN gate PersonalityGate to automatically extract features from the information cascade process for personality recognition and then utilize obtained personality traits to enhance cascade prediction iteratively. Experiments on two real-world datasets demonstrate the effectiveness of our proposed PersonalityGate and MeCAPE framework for both cascade prediction and personality recognition. In future work, we plan to improve the flexibility of  PersonalityGate with an inductive paradigm and apply it to a wider range of GNN-based tasks such as network alignment other than cascade prediction.

\ifCLASSOPTIONcompsoc
  \section*{Acknowledgments}
\else
  \section*{Acknowledgment}
\fi

This work is supported by the National Natural Science Foundation of China (Grant No. 61872002, U1936220), the University Natural Science Research Project of Anhui Province (Grant No. KJ2019A0037), the National Key R$\&$D Program of China (Grant No.2019YFB1704101) and the Anhui Provincial Natural Science Foundation (Grant No. 2008085QF307).

\ifCLASSOPTIONcaptionsoff
  \newpage
\fi



%
\bibliographystyle{unsrt}
\bibliography{PersonalityGate-ref}



%

\begin{IEEEbiography}[{\includegraphics[width=1in,height=1.25in,clip,keepaspectratio]{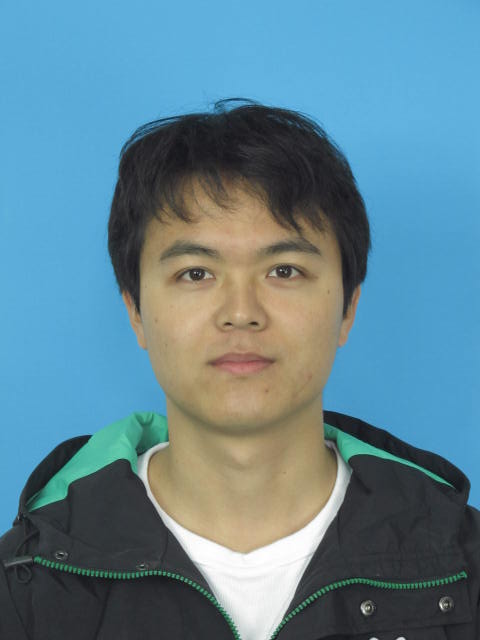}}]{Dengcheng Yan}
received the B.S. and Ph.D. degrees from the University of Science and Technology of China, in 2011 and 2017, respectively. From 2017 to 2018, he was a Core Technology Researcher and a Big Data Engineer with Research Institute of Big Data, iFlytek Company Ltd. He is currently a Lecturer with Anhui University, China. His research interests include software engineering, recommendation systems and complex networks.
\end{IEEEbiography}

\begin{IEEEbiography}[{\includegraphics[width=1in,height=1.25in,clip,keepaspectratio]{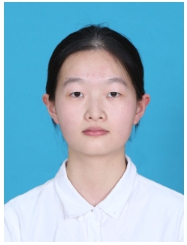}}]{Jie Cao}
received the B.S. degree from Anhui University, China, in 2019. She is currently pursuing the M.S degree with Anhui University of computer science. Her research interest includes graph neural network and personality prediction.
\end{IEEEbiography}

\begin{IEEEbiography}[{\includegraphics[width=1in,height=1.25in,clip,keepaspectratio]{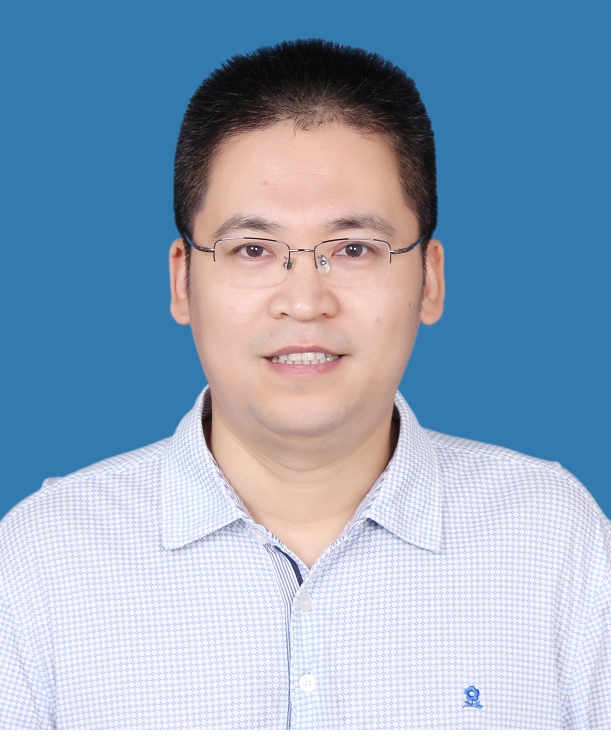}}]{Yiwen Zhang}
received the Ph.D. degree in management science and engineering from the Hefei University of Technology, in 2013. He is currently a Professor with the School of Computer Science and Technology, Anhui University. His research interests include service computing, cloud computing, and big data analytics.
\end{IEEEbiography}

\begin{IEEEbiography}[{\includegraphics[width=1in,height=1.25in,clip,keepaspectratio]{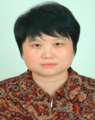}}]{Hong Zhong}
received the Ph.D. degree in computer science from the University of Science and Technology of China in 2005. She is currently a Professor and a Ph.D. Supervisor with the School of Computer Science and Technology, Anhui University. She has over 120 scientific publications in reputable journals, academic books, and international conferences. Her research interests include applied cryptography, IoT security, vehicular ad hoc networks, cloud computing security, and software-defined networking (SDN).
\end{IEEEbiography}





\end{document}